# Stochasticity Invariance Control in $Pr_{1-x}Ca_xMnO_3$ RRAM to enable Large-Scale Stochastic Recurrent Neural Networks

**Vivek Saraswat (svivek@ee.iitb.ac.in) and Udayan Ganguly (udayan@ee.iitb.ac.in)**
Department of Electrical Engineering, Indian Institute of Technology Bombay, Mumbai-400076, India

**Abstract**

Emerging non-volatile memories have been proposed for a wide range of applications from easing the von-Neumann bottleneck to neuromorphic applications. Specifically, scalable RRAMs based on $Pr_{1-x}Ca_xMnO_3$ (PCMO) exhibit analog switching have been demonstrated as an integrating neuron, an analog synapse, and a voltage-controlled oscillator. More recently, the inherent stochasticity of memristors has been proposed for efficient hardware implementations of Boltzmann Machines. However, as the problem size scales, the number of neurons increase and controlling the stochastic distribution tightly over many iterations is necessary. This requires parametric control over stochasticity. Here, we characterize the stochastic Set in PCMO RRAMs. We identify that the Set time distribution depends on the internal state of the device (i.e., resistance) in addition to external input (i.e., voltage pulse). This requires the confluence of contradictory properties like stochastic switching as well as deterministic state control in the same device. Unlike, 'stochastic-everywhere' filamentary memristors, in PCMO RRAMs, we leverage the (i) stochastic Set in negative polarity and (ii) deterministic analog Reset in positive polarity to demonstrate 100× reduced Set time distribution drift. The impact on Boltzmann Machines' performance is analyzed and as opposed to the "fixed external input stochasticity", the "state-monitored stochasticity" can solve problems 20× larger in size. State monitoring also tunes out the device-to-device variability effect on distributions providing 10× better performance. In addition to the physical insights, this study establishes the use of experimental stochasticity in PCMO RRAMs in stochastic recurrent neural networks reliably over many iterations.

**Introduction**

Emerging non-volatile memories like RRAMs, MRAMs and FeRAMs have generated immense interest and excitement in the computing community in the past couple of decades. Particularly, the versatility of their applications ranging from conventional dense memory arrays to neuro-synaptic characteristics for neuromorphic hardware and algorithms has been highlighted[1–5]. Memristors have been experimented with the most, owing to their simple capacitor-like structure. Filamentary RRAMs like $HfO_x$ RRAMs have been adopted widely due to their CMOS compatibility and fast switching[6–8]. The non-filamentary switching RRAMs, on the other hand, like those based on $Pr_{1-x}Ca_xMnO_3$ material system are popular for low device-to-device variability, scalable currents, analog switching states, and volatile and non-volatile switching modes[9–13]. As a result of these properties, they have been demonstrated for a wide range of applications ranging from analog synapses[14], integrating neurons[10] to voltage-controlled oscillators.

Another property common to both filamentary and non-filamentary RRAMs is the presence of inherent stochasticity or cycle-to-cycle variability in switching. These devices are natural noise generators which has found numerous applications from security to stochastic neural networks for optimisation[15,16]. A random number generation is a task that the noise-free, deterministic, digital computer is not good at[17]. As a result, optimization networks, like the Boltzmann Machines, which run for millions of iterations of stochastic dot product for a problem of reasonable size are not well-suited for a conventional computing hardware[18]. In this scenario, memristors enable promising solutions. First, dedicated architectures with memristor crossbar arrays have been proposed to speed up the vector matrix multiplication[18]. Gradual state control in memristors enable scaled synapses with analog memory[13,19,20]. Second, the concept of memristor based stochastic neurons is also explored extensively with switching stochasticity or Read stochasticity being proposed for binary probabilistically switching neurons[15,16,21–23] (Fig. 1. (a)). Stochasticity is important because it helps escape local minima when performing gradient descent on an optimization network. More and more iterations of stochastic gradient descent are required as the problem size rises to convincingly explore the wide solution space (Fig. 1. (b)). Naturally, longer term control over stochastic switching distribution parameters is crucial to solve different types of optimization problems.

The requirements from a stochastic neuron are summarized in Fig. 1. (c). Although conceptual demonstrations of controllable stochasticity have been achieved in different material systems[16,24,25], the consistency to generate the stochastic distribution repeatedly over many iterations is still lacking. For practical problems of large size, presence of controllable stochasticity for a short duration is not sufficient. A guarantee of a particular stochastic distribution independent of how long the circuit has been running is crucial. Further, an optimization problem of large size equips a large number of stochastic neurons and hence multiple RRAM devices. The device-to-device variations in the stochastic distributions is another aspect which affects practical implementations of Boltzmann Machines.

In this paper, we demonstrate control of stochastic distribution over an extended duration through a confluence of both internal state (i.e., resistance) and external input (i.e., voltage pulse). First, the Set time stochasticity of PCMO RRAM is measured as a function of the initial high resistance state of the device to establish causal dependence. Second, the stochastic distribution parameters, defined by internal state and external input, are compared to fixed "electrical inputs-only" characterization to show reduced distribution drift. Third, the gradual, analog, deterministic Reset process enables excellent internal state control. Finally, the reduced distribution drift significantly increases the size of optimization problems that can be solved reliably – without added complexity of repeated distribution re-calibrations. In addition, the internal state monitoring significantly tunes out the negative impact of the device-to-device variations by aligning the stochastic distributions across devices.

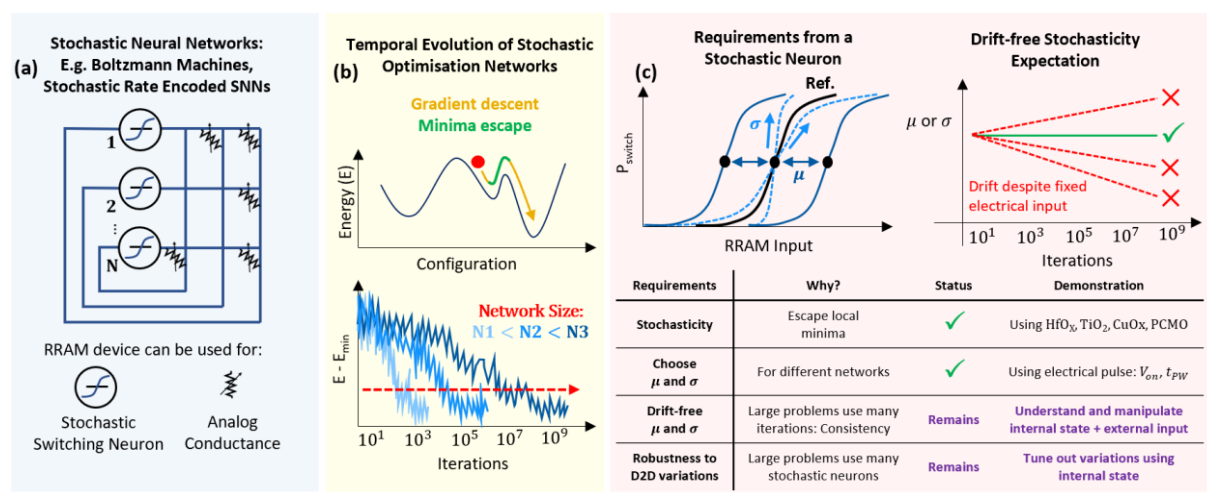

Fig. 1. **Setting up requirements from a stochastic neuron:** (a) Utility of RRAM devices in different stochastic neural network components, (b) Energy evolution using gradient descent and stochastic minima escape over iterations for different size problems, (c) Requirements from a stochastic neuron and their status: Presence of stochasticity, Choosing stochastic distribution parameters, Consistency of that choice over multiple stochastic iterations, Robustness to device-to-device variations.

The dependence of stochastic distribution parameters on the internal state of the device, although guarantees consistent stochastic distributions, presents a new challenge. The requirement, now, from a "stochastic" device, is to achieve a "deterministic" state to begin with.

From a system's perspective, deterministic analog synapses are highly attractive for compact multiply and accumulate operation in cross-bar arrays[1]. Further, we hypothesize that drift-free controllable stochasticity in Set would require the control of the initial resistance state in a deterministic Reset. The choice of memristors for a system is as follows. First, a deterministic, gradual RRAM with good internal state control for analog synaptic application but would provide no harvestable stochasticity (Fig. 2(a)). Second, filamentary RRAMs typically have a lot of variability and stochasticity in all regions of operation to produce binary and hence bulky synapses[26,27] (Fig. 2(b)). Further, the stochastic switching distribution is not controllable as the initial and final states are not well controlled. Thus, a confluence of deterministic Reset to control the initial resistance followed by stochastic Set is essential for drift-free controllable stochastic distribution. Our proposal is to use PCMO RRAM which serendipitously combines these contradictory properties. To elaborate, PCMO RRAMs perform a gradual and analog deterministic Reset in the positive polarity along with an abrupt, stochastic Set in the negative polarity. Thus, all the ingredients of a stochastic neuron are enabled through polarity control suitable for practical realizations of Boltzmann Machine networks (Fig. 2(c)). The gradual deterministic Reset also enables compact, analog memory for synaptic cross-bar array. Thus, PCMO RRAM is used to investigate the realization of our proposal experimentally.

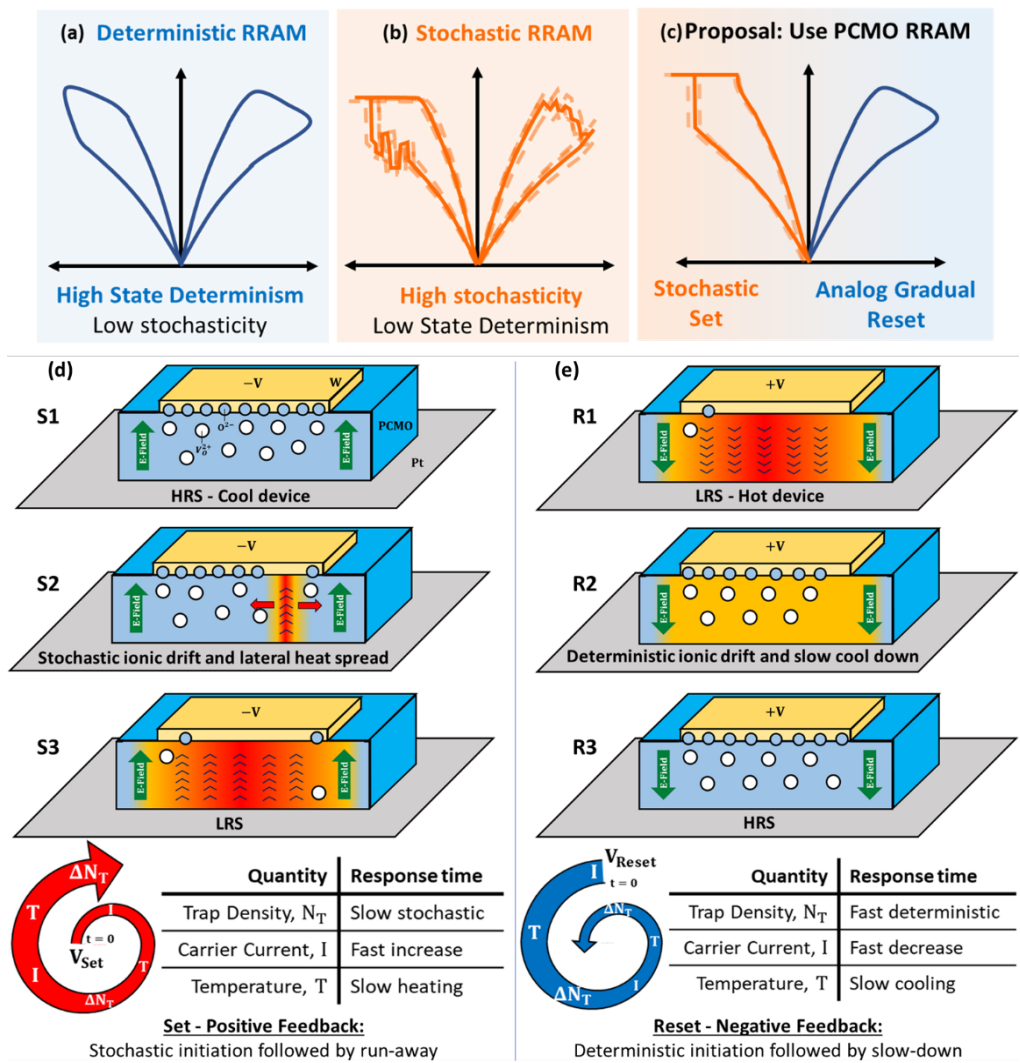

Fig. 2. **Proposed benefits of PCMO RRAM as a stochastic neuron:** PCMO RRAM combines the features of (a) deterministic RRAM and (b) stochastic RRAM different polarities to demonstrate (c) stochastic Set and analog gradual Reset. (d) Set and (e) Reset mechanisms for PCMO RRAM. Set process starts as a stochastic ionic drift followed by thermal run-away (positive feedback), Reset starts as a deterministic ionic drift followed by slow cooldown (negative feedback)

The switching mechanism for the Set and the Reset processes in PCMO RRAM are shown in Fig. 2 (d) and (e) and discussed extensively before[28]. The Set process starts out as a stochastic ionic (or vacancy) motion followed by positive feedback between carrier current, self-heating and further ionic motion. On the other hand, the Reset process starts out with deterministic ionic (or vacancy) motion followed by negative feedback between temperature cool-down and further ionic motion. The effect of this asymmetry on amount of stochasticity in Set vs Reset switching is discussed in detail in this work.

**Experimental Section: Characterization Methods**

A. Set Stochasticity

PCMO RRAM device has a simple metal-oxide-metal structure: a thin layer (60 nm) of PCMO sandwiched between a top W (10 μm × 10 μm) and bottom Pt electrode on a Si/SiO$_2$/Ti substrate[29] (Fig. 3(a)). The structure, fabrication and characterization tool details are discussed in Supplementary Section S1. The PCMO device resets into a high resistance state (HRS) in positive polarity and sets into a low resistance state (LRS) in the negative polarity. The current in PCMO RRAMs is governed by bulk trap-modulated space charge limited conduction (SCLC). The switching mechanism is attributed to the reaction-drift of oxygen vacancies to and from the W/PCMO interface and the PCMO bulk[28]. The mean behavior of short (~ ns) and long-time range (~ s) current transients have been modelled and published using the reaction-drift and thermodynamic models[28,30].

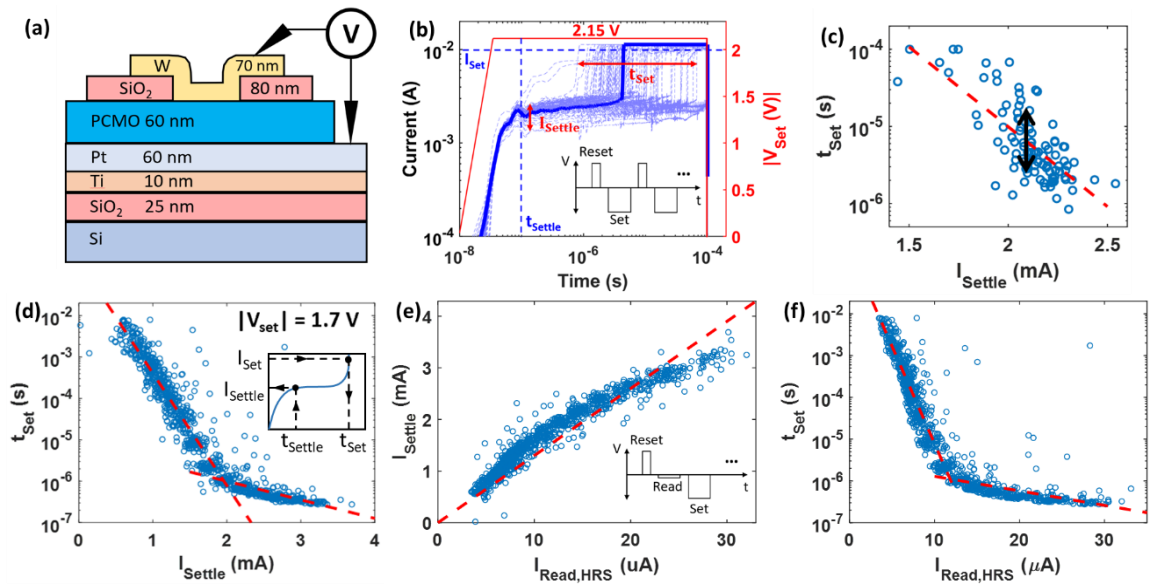

Fig. 3. **Experimentally measured stochasticity:** (a) Experimental Set current transients repeated 100 times for fixed $V_{Reset}$, fixed $V_{Set}$ scheme (inset). The solid blue curve is a typical Set transient. The lighter blue transients show cycle to cycle variations. (b) Scatter of $t_{Set}$ vs $I_{Settle}$ (indicating monotonic decrease in $t_{Set}$ with $I_{Settle}$, hence the red dashed trendline). The spread in $t_{Set}$ at a given $I_{Settle}$ is marked using a double arrow black line. (d) $t_{Set}$ vs $I_{Settle}$ for $t_{Settle}$ = 100 ns, $I_{Set}$ = 10 mA (inset) showing two slope characteristics in log-linear scale (dashed red line), (e) $I_{Settle}$ vs $I_{Read,HRS}$ enabled by Reading the HRS current before Set (inset) showing linear relationship (dashed red line), (f) $t_{Set}$ vs $I_{Read,HRS}$ showing two-slope characteristics in log-linear scale (dashed red line)

Fig. 3. (b) shows the experimental Write current transients for the Set pulse. When the Set voltage is applied, the current rises and settles to an initial level. At this timescale, the drift of vacancies is too slow to produce any further change, so the current stays almost flat. At a later timescale, when the drift timescale of vacancies matches with the measurement timescale, a sharp increase in the current to compliance is observed indicating abrupt Set in the logarithmic time axis[28]. The Set current crosses a threshold $I_{Set}$ to define a Set time, $t_{Set}$. The abrupt shoot-up and the well-defined Set time are explained by a positive feedback between vacancy motion, current, and self-heating induced

temperature to create a run-away process limited by compliance. This Set process is repeated several times as a part of the following scheme: a fixed Set voltage pulse followed by a Reset pulse of fixed voltage and pulse width (inset of Fig. 3(b)). These cycles of Reset and Set gives rise to a family of Set transients in the device as shown in Fig. 3(b). These transients denote the stochastic Set in a PCMO RRAM. The distribution in Set times can be used to generate a stochastic neuron for Boltzmann Machines[16].

A closer inspection of the Set transients stochasticity reveals that there is a distribution in the initial currents, $I_{Settle}$ measured at a pre-defined $t_{Settle}$ after the input voltage rise is complete. This indicates that the present distribution in $t_{Set}$ may not be entirely stochastic or based only on the applied Set voltage. The dependence of the $t_{Set}$ on the $I_{Settle}$ is shown in Fig. 3(c). Clearly the total $t_{Set}$ scatter has a deterministic component related to the $I_{Settle}$ and a stochastic component at any given $I_{Settle}$. However, the $I_{Settle}$ should indeed be a function of the initial HRS of the device. We test this next.

## B. HRS Dependence

In order to observe the dependence of $t_{Set}$ on the initial high resistance state (HRS), we repeat the Reset, Set measurements as in previous section but add a Read pulse before the Set. We perform this experiment for a fixed $|V_{Set}| = 1.7$ V. The Reset voltages are incremented or decremented per cycle in the range of [1.5 – 2.5] V so as to be able to sweep a large range of initial HRS. The initial HRS is measured using the Read current, $I_{Read,HRS}$ using a fixed small voltage of -0.5 V. The $t_{Set}$ and $I_{Settle}$ are extracted from the Set transients as mentioned before.

First, we observe that the Set transients reveal a dependence of $t_{Set}$ on $I_{Settle}$ (Fig. 3(d))). It is a two-slope dependence in the log-linear domain indicating an exponential decrease in $t_{Set}$ as the $I_{Settle}$ increase followed by a saturation in $t_{Set}$ for even higher $I_{Settle}$. This limit indicates the fastest switching times (~ 100 ns) for the given Set voltage. With the help of the Read pulse, we can now demonstrate that the $I_{Settle}$ is indeed linearly related to the $I_{Read,HRS}$ (Fig. 3(e)). Hence, the dependence that we are observing in $t_{Set}$ can ultimately be mapped to this initial HRS of the device (Fig. 3(f)). Again, as the Read current rises, the $t_{Set}$ falls exponentially followed by saturation for even higher $I_{Read,HRS}$. Earlier studies have shown a similar behavior of $t_{Set}$ with direct input of Set voltage. This is a *first demonstration* of the $t_{Set}$ dependence on the internal state i.e., initial HRS before Set of the PCMO RRAM device.

## C. Controllable Stochasticity

It can be demonstrated that the stochastic cycle-to-cycle distribution of $t_{Set}$ in PCMO RRAM devices is lognormal at any given HRS (details in Supplementary Section S2). The mean dependence of $t_{Set}$ on Set voltage, $|V_{Set}|$ is well known and shown earlier[28,31]. We add the $t_{Set}$ vs $HRS$ characterization discussed in previous sections for a wide range of $|V_{Set}|$ (1.6 – 2.2 V) values where $HRS = V_{Read}/I_{Read,HRS}$ and $V_{Read} = -0.5$ V. The experimental $t_{Set}$ is plotted in Fig. 4(a). We observe that the mean $t_{Set}$ increases as the HRS increases or the $|V_{Set}|$ decreases. We fit a surface through the experimental data of $\log t_{Set}$ with quadratic dependencies on $|V_{Set}|$ and HRS. The surface is an excellent fit with 0.9 R-squared metric. This is plotted in Fig. 4(b) and expressed as the mean of $\log t_{Set}$. Next, we calculate the variation of experimental data with respect to this mean and obtain the standard deviation of $\log t_{Set}$ plotted in Fig. 4(c). This enables the complete characterization of $t_{Set}$ distribution with the lognormal parameters $\mu$ and $\sigma$ as a function of the $|V_{Set}|$ and HRS.

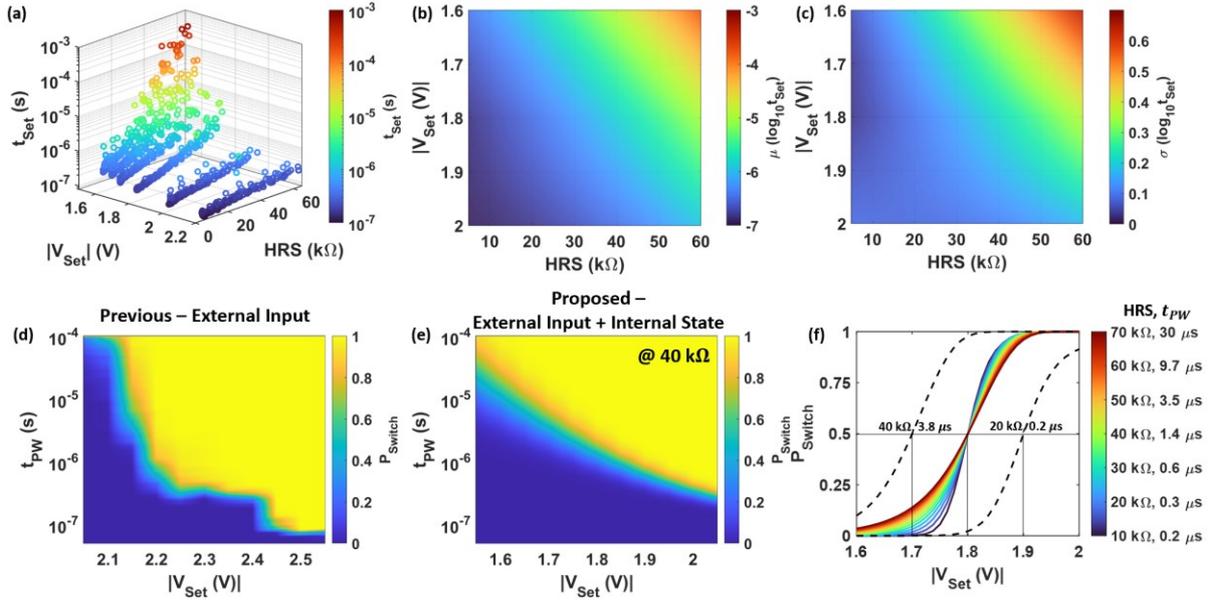

Fig. 4. **Controllable Stochasticity:** (a) Experimental stochastic $t_{Set}$ measured for a range of HRS for different $|V_{Set}|$, Bi-quadratic surface fit (R-squared ~ 0.9) is used for (b) Extracted $\mu$ of $\log_{10} t_{Set}$, (c) Extracted $\sigma$ of $\log_{10} t_{Set}$ assuming lognormal distribution of $t_{Set}$ at each ($|V_{Set}|$, HRS), (d) $P_{Switch}$ extracted from previous methodology of fixed electrical input compared with (e) $P_{Switch}$ extracted from proposed methodology of fixed electrical input and internal state of the device, (f) $P_{Switch}$ vs $|V_{Set}|$ demonstrating control of bias point and steepness of switching probability as a function of internal state (HRS) and $t_{PW}$.

Now, we can characterize the stochastic switching neuron. A neuron is said to switch if the $t_{Set}$ (i.e., time to reach $I_{Set}$) is lower than the applied pulse width $t_{pw}$. Since the $t_{Set}$ is a lognormal random variable dependent on $|V_{Set}|$ and HRS, the switching is stochastic and results in a stochastic neuron. The switching probability can be obtained as:

$$P_{Switch} = P(t_{Set} \leq t_{PW}) \text{ where } t_{Set} \text{ is lognormal}$$

Given the external inputs $|V_{Set}|$, $t_{pw}$ and internal state HRS, the lognormal distribution parameters for $t_{Set}$ can be found ($\mu$(HRS, $|V_{Set}|$) and $\sigma$(HRS, $|V_{Set}|$)) and are plotted in Fig. 4(b),(c) followed by calculating the cumulative probability distribution (CDF) function at $t_{pw}$. Thus $P_{Switch}$ is a function of $|V_{Set}|$, HRS and $t_{pw}$. The switching probability of the device was studied earlier as a function of $|V_{Set}|$ and $t_{pw}$ alone without monitoring the HRS[16]. The results are plotted in Fig. 4(d). In this work, we calculate the switching probability as a function of $|V_{Set}|$ and $t_{pw}$ at a particular HRS (40 kΩ) as shown in Fig. 4(e). Clearly, capturing the HRS dependence has improved the behavior of the $P_{Switch}$. It is now well defined and predictable with respect to electrical inputs.

A stochastic neuron is typically used with a fixed $t_{pw}$ while $|V_{Set}|$ serves as the input to control $P_{Switch}$. We plot the $P_{Switch}$ vs $|V_{Set}|$ for a fixed HRS and $t_{pw}$ to get sigmoid-like characteristics (Fig. 4(f)). In order to center the sigmoid around a particular $|V_{Set}|$ = 1.8 V, we choose $t_{pw}$ as the mean of the lognormal $t_{Set}$ for the given $|V_{Set}|$ and HRS. This ensures that $P_{Switch}$ (1.8 V) = 0.5 (since CDF at mean is 0.5). The center of the sigmoid can be shifted to any voltage of choice by this methodology (as seen by black dashed lines in Fig. 4(f)). Next, we vary the HRS (and correspondingly the $\log t_{pw} = \mu$ of $\log t_{Set}$ at 1.8V and chosen HRS such that $P_{Switch}$ (1.8 V) = 0.5 is maintained) to get sigmoids of varying steepness. Thus, a stochastic neuron with controllable center and steepness of sigmoidal switching probability is demonstrated.

## D. Drift – free Stochasticity

Drift-free stochasticity means the ability to generate instances from the exact same stochastic distribution over time over multiple iterations. In other words, it is the ability to control all the inputs/parameters external and internal to the device that determine the stochastic distribution parameters (mean and standard deviation). We have identified Set voltage and HRS as the inputs that affect $t_{Set}$ and its stochasticity. In this experiment, we observe the distribution of $t_{Set}$ along multiple cycles while measuring the HRS before each Set (Fig. 5(a)). Every cycle is an application of a Reset, Read and Set pulse. The Reset voltage of consecutive cycles is decreased in small amounts (~ 0.05 V) which leads to progressively lower HRS points. Once we achieve a low enough LRS, the direction of change of Reset voltage in the consecutive cycles is reversed which leads to progressively higher HRS points. This back and forth sweeping of the HRS range and the resultant $t_{Set}$ points are measured for 1000 cycles (Fig. 5(a)). The dependence of $t_{Set}$ on HRS is fairly static over cycles (evident by HRS color bands parallel to the x-axis) indicating complete control over $t_{Set}$ distribution using internal state HRS and external input $V_{Set}$. The variation of $t_{Set}$ as a function of the mean $t_{Set}$ is plotted in Fig. 5(b) to indicate that the $t_{Set}$ although controllable is indeed stochastic.

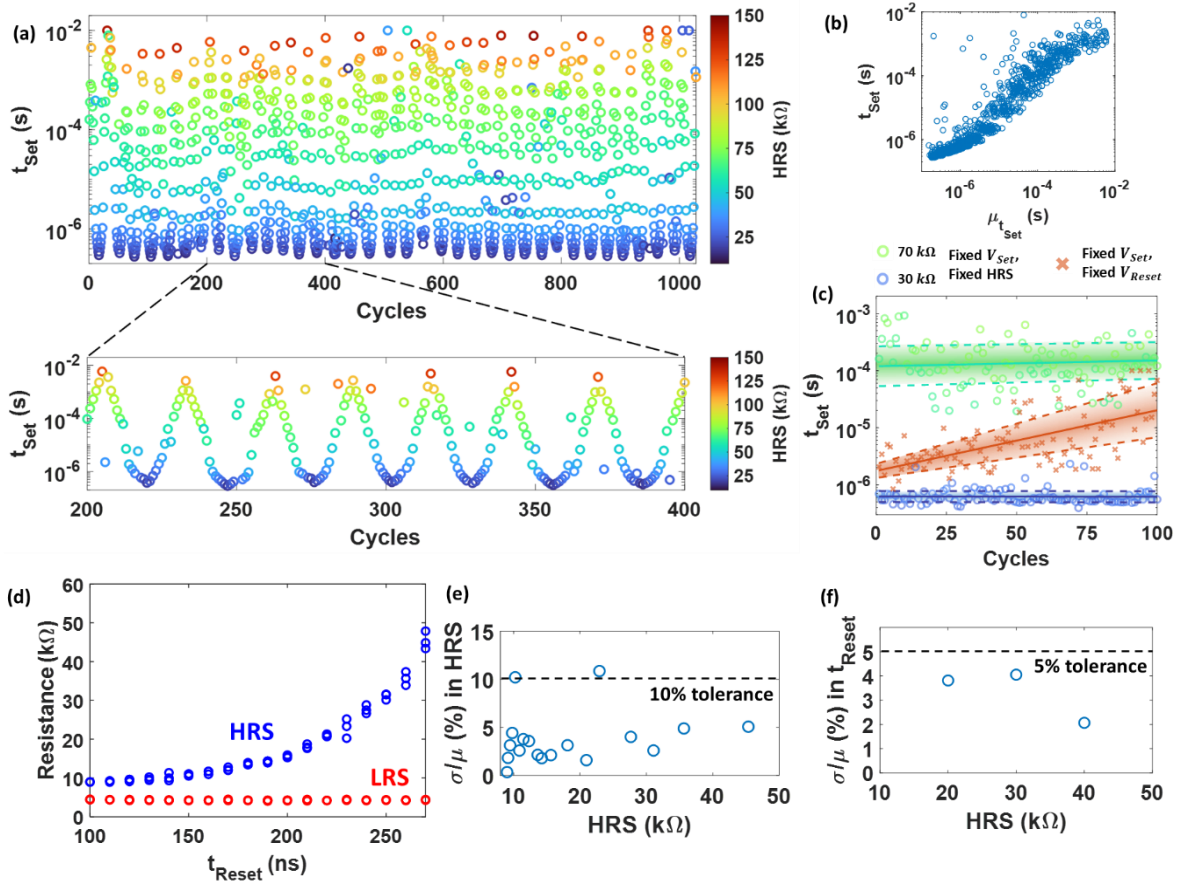

Fig. 5. **Demonstration of consistent stochasticity:** (a) Repeatable control over $t_{Set}$ over 1000 Reset-Read-Set cycles, with fixed $V_{Set}$ and Read pulse enabled HRS measurement (color), $t_{Set}$ vs cycles zoomed in for 200 cycles to show repeatability of distribution, (b) experimental $t_{Set}$ vs the mean $t_{Set}$ to show presence of stochasticity over 1000 cycles, (c) $t_{Set}$ vs Cycles for fixed $V_{Set}$, fixed HRS ($\pm$ 10% tolerance) cycling (circles) vs fixed electrical input ($V_{Set}$ and $V_{Reset}$) cycling (crosses) (markers – experimental, solid line – mean, dashed line – sigma). (d) Reset Resistance Transient at $V_{Reset}$ = 1.9 V as a function of Reset pulse widths for 3 sweeps demonstrating high precision, analog and gradual control over HRS using the Reset operation starting from a fixed LRS ($|V_{Set}|$ = 3 V, $I_{comp}$ = 10 mA), (e) HRS can be controlled within a tolerance of 10% variation using Reset, (f) $t_{Reset}$ variability is lower than 5% for different HRS values.

The comparison of control over $t_{Set}$ distribution with and without monitoring the internal state is shown in Fig. 5(c). If we perform Reset-Set cycles of fixed magnitude and pulse duration without actively monitoring the internal state (as in Fig. 3(b)), the $t_{Set}$ distribution has the potential to drift a lot (cross markers in Fig. 5(c)). The drift is 1 decade in mean and 0.34 decades in standard deviation of the lognormal $t_{Set}$ over 100 cycles. On the other hand, if the HRS is ensured to be within 10% tolerance of a target over multiple cycles, the control over $t_{Set}$ is drastically improved (circle markers in Fig. 5(c)). We show this for two bounding HRS values. The resultant drift now is reduced significantly to 0.01 decades in mean and 0.003 decades in standard deviation. The resultant drift is so minimized that it is statistically insignificant and points to a simpler model – **With HRS and $V_{Set}$ fixed, $t_{Set}$ distribution is fully determined across iterations and time** (unless of course if something fundamentally ages about device – reliability concern – however switching endurance is well-demonstrated earlier[11]). This result is extremely significant for use of these devices to reliably and repeatedly generate random numbers from a fixed distribution. For use in a Boltzmann Machine, the number of iterations before the network converges can easily run into millions of cycles depending on the network size. This means every stochastic neuron in the network undergoes large number of samplings during the course of network evolution (exactly one neuron being sampled every iteration). We will analyze the impact of drifting stochastic distribution in next sections.

### E. Resistance state controllability in PCMO RRAM

As evident from the previous section, the ability to consistently produce identical stochasticity over multiple Reset-Set cycles depends on the ability to control the internal state HRS with precision. This is where PCMO RRAM devices are particularly attractive. Not only do they show Set based stochasticity which is controllable but they are also excellent analog memories with gradual controllable resistance using the opposite polarity Reset[11,20,30,31]. This difference arises out of the fundamentally different feedback mechanisms at operations in Set (positive feedback, abrupt) and Reset (negative feedback, gradual) discussed extensively in the reaction drift model proposed for the switching phenomena in these memories[28]. We perform Reset resistance transient measurements where starting from a fixed low resistance state (LRS) a Reset pulse of fixed magnitude (1.8 V) and pulse width is applied (Fig. 5(d)). A fixed LRS is ensured by choosing a high $|V_{Set}|$ = 3 V and compliance current, $I_{comp}$ = 10 mA. As the pulse width of the Reset pulse is increased, the Read resistance after the Reset pulse starting from a fixed LRS gradually increases (Fig. 5(d)). The pulse width sweep is performed multiple times back and forth (3 times) to show precise control over resistance using Reset over a wide range of resistances. All resistance measurements have a sub-10% variation with a very small fraction exceeding even 5% variation (Fig. 5(e)). In order to demonstrate the asymmetry in stochasticity, the variation in $t_{Reset}$ is plotted for three different HRS values ($\pm$ 20% tolerance) (Fig. 5(f)). Compared to lognormal distributions of $t_{Set}$ with $\sigma(\log_{10} t_{Set})$ varying from 0.1 to 0.7 decades (Fig. 4(c)), the distribution of $t_{Reset}$ is linear with sub-5% variation. This demonstrates precise HRS control using Reset pulses in otherwise Set based stochastic PCMO RRAM. This behavior is very powerful since resistance controllability makes them the choice of memories for the crossbar array weights in a Boltzmann Machine along with stochastic neurons on the edge of this crossbar utilizing the Set stochasticity. A single material systems-based Boltzmann Machines chip is realizable using PCMO RRAM devices.

# Network Simulation Section: Performance of Boltzmann Machines

## A. Impact of cycle-to-cycle drift:

The utility of RRAM devices in efficient implementation of Boltzmann Machines has been explored extensively[18]. The possibility of analog multiply and accumulate in the crossbar array and the generation of analog approximate sigmoid switching probability using inherent switching stochasticity provide significant hardware implementation benefits[16]. However, the Boltzmann Machines are networks which run for a large number of iterations performing stochastic gradient descent before converging. Hence, demonstration of controllable stochasticity is necessary but not sufficient. It is well known that PCMO RRAMs have good endurance with each device capable of switching tens of thousands of cycles[11]. However, it is extremely essential to be able to reproduce the switching probability of a neuron for this large number of Reset-Set cycles. As seen in previous section (Fig. 5(c)), if the internal state of the device drifts over these cycles, the stochastic distributions drift as well. This will result in changing $P_{Switch}$ characteristics which can potentially affect the performance of Boltzmann Machines. In this section, we analyse the impact of HRS drift over cycles on the maximum size of the Max-Cut optimisation problem that a Boltzmann Machine, built out of these devices, can solve.

The Max-Cut problem requires us to find a bi-partition of a graph with N nodes and E edges such the sum of the weights of the edges crossing between these partitions is maximized. The input to this problem is the adjacency matrix of weights $w_{ij}$ connecting any two nodes $i$ and $j$. Let us represent the current configuration of the network by the vector $x$ of nodes where $x_i$ is 0 if it belongs to first partition and it is 1 if it belongs to second partition. The quantity to be maximized then can be expressed as[16]:

$$M(x) = \Sigma_{i=1}^{N} \Sigma_{j=i+1}^{N} w_{ij} \cdot [(1-x_i).x_j + (1-x_j).x_i]$$

The energy to be minimized is then $E(x) = -M(x)$:

$$E(x) = b^T x - \frac{1}{2} x^T W_B x$$

Where $b_i = -\Sigma_{j=1}^{N} w_{ij}$ and $W_{B,ij} = -2w_{ij}$

The energy is expressed in the standard form for the Boltzmann Machine. The machine performs stochastic gradient descent if at every iteration exactly one neuron switches with the sigmoidal probability as a function of $u_i = \Sigma_j W_{B,ij} x_j - b_i$.

The Max-Cut instances are chosen from standard benchmark libraries for this work[32,33]. These are random or planar graphs of size [100 – 3000] nodes with weights chosen from {-1,0,1}. The sigmoidal switching probability is achieved using the RRAM switching probability after scaling and shifting the input $u_i$ to the range of input voltages that RRAM responds to (Fig. 4(f)). The ideal Boltzmann Machine solution using a fixed switching probability characteristics for a 125 nodes problem is shown in Fig. 6(a). The energy reduces stochastically to reach the best-known solution over $\sim 10^4$ iterations. In each iteration, exactly one RRAM device undergoes stochastic switching. The corresponding separation of the desired bi-partition as energy reduces is shown on the left in Fig. 6(a). Multiple stochastic runs are performed to display the variation in the stochastic performance of the Boltzmann Machine.

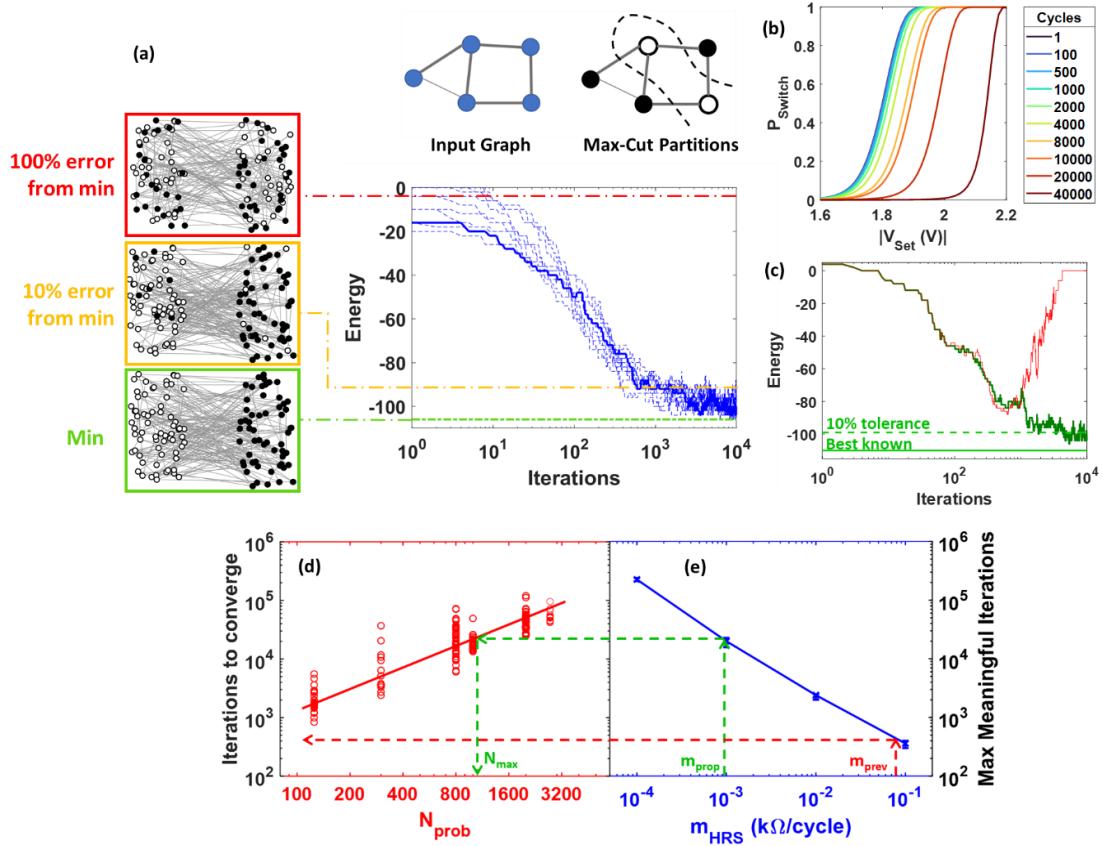

Fig. 6. **Ideal Solution to Max-Cut Problem and Effect of cycle-to-cycle HRS drift:** (a) The objective is to find a bi-partition of the input graph maximizing the weights in the cut. Hollow circles and filled circles denote the desired bi-partition. As energy evolves over iterations, the partition separates out into the desired configuration, (b) Switching probability characteristics of a single neuron which has undergone different number of reset cycles assuming a constant HRS drift per Reset, (c) Energy vs iterations of ideal stochastic gradient descent ($m_{HRS} = 0$) and in the presence of HRS drift ($m_{HRS} = 0.01$ k$\Omega$/cycle). (d) Iterations taken to converge within 10% of best-known solution for different problem sizes (multiple runs and multiple problem instances ($\sim 25$ runs) of each problem size), (e) maximum meaningful iterations of stochastic gradient descent for a given extent of HRS drift

For an N node problem, we require N stochastic neurons or N RRAM devices. Every iteration, exactly one RRAM device is Reset-Set to evolve the configuration. Every Reset-Set causes the HRS to drift for that particular RRAM device. The effect of HRS drift on the switching characteristics for different number of Reset-Set cycles undergone is shown in Fig. 6 (b). The drift is modelled as a linear shift with slope $m$ (in k$\Omega$/cycle). There is significant shift in the sigmoid curves over cycles. The effect of HRS drift is now observed on the energy transient of the stochastic gradient descent in Fig. 6(c). Compared to the ideal case (m = 0) which takes some iterations to converge (i.e., reach within 10% of the best-known solution of the problem), the HRS drift ($m_{HRS}$ = 0.01 $k\Omega$/cycle) case performs meaningful stochastic gradient descent up to only some max no. of iterations after which it is no longer moving towards the optimal energy. We can calculate the typical number of iterations required to solve the Max-Cut problem using an ideal no HRS drift neuron for different problem sizes (Fig. 6(d)). We can also calculate the maximum number of meaningful iterations of stochastic gradient descent that a network can perform with neurons of a given HRS drift (Fig. 6(e)). For any given $m_{HRS}$, we can thus find the maximum size of problem that can be solved. With the improved consistency in stochasticity obtained using the fixed HRS method in PCMO RRAM results in an 20X improvement (sub-100 to above-1000) in the size of the problem that can be solved. Hence to solve practical problems of large size, drift-free stochasticity is key and essential for efficient hardware implementations of Boltzmann Machines.

## B. Impact of device-to-device variations:

Large problems also make use of large number of stochastic neurons and hence many stochastic RRAM devices. These devices vary from each other in terms of their stochastic $t_{Set}$ distribution. Fig. 7(a) shows the $t_{Set}$ vs HRS scatter for 6 different devices at the same $|V_{Set}|$ = 1.9 V. If the $t_{Set}$ distribution for the 6 devices at a given HRS of 140 $k\Omega$ (with $\pm$ 20% variation in HRS) is plotted, the distributions are widely separated from each other with the primary difference in their $\mu(\log_{10} t_{Set})$ – about 20% variation device-to-device (Fig. 7(b)). This can potentially affect the quality of the solution obtained by Boltzmann Machine. Once again, the HRS controllability in PCMO RRAM devices helps to tune out this variability once before the network evolution begins. The HRS of every device is set (with realistic precision of 20% variation as shown in Fig. 7(b)) to the value that corresponds to a fixed $\mu(\log_{10} t_{Set}) = -5$. The new $t_{Set}$ distributions using the HRS controlled scheme are plotted again in Fig. 7(c). The device-to-device variation in stochastic distributions is brought down to just 2% (as opposed to 20% in the fixed HRS method).

The effect of device-to-device variations in $t_{Set}$ translates to variation in switching probability curves of different devices (Fig. 7(d)). Different $P_{switch}$ curves for each stochastic neuron affect the minimum settling energy of the Boltzmann Machines (Fig. 7(e)). Higher variations result in a higher settling energy and hence more error % with the optimal energy. This error % in settling energy with respect to the ideal settling energy is plotted for different amounts of device-to-device variability (Fig. 7(f)). Compared to 50% error from ideal for 20% device-to-device variations, the HRS controllability (reduced variations to 2%) allows the error to improve 10× to just 5%. Thus, the knowledge of stochasticity affecting parameters namely HRS and $V_{Set}$ and the ability to control HRS in the same device helps to tackle practical problems of cycle-to-cycle drift and device-to-device variations and enable large scale stochastic neural networks.

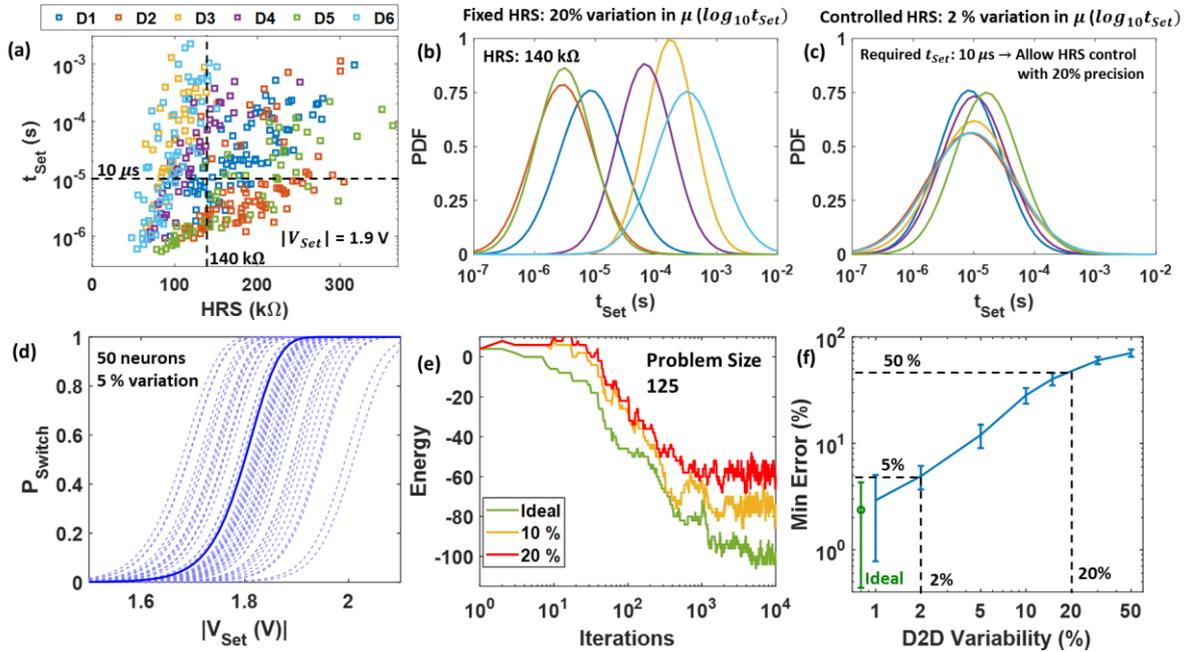

Fig. 7. **Effect of device-to-device variations:** (a) Experimental $t_{Set}$ vs HRS for 6 different devices at $|V_{Set}|$ = 1.9 V, (b) probability distribution function (PDF) of the six devices at a fixed HRS of 140 k$\Omega$ ($\pm$ 20% precision) shows $\mu(\log_{10} t_{Set})$ variability of 20% device-to-device, (c) Improved device-to-device variability to 2% by controlling HRS of each device (with $\pm$ 20% precision) to ensure that $\mu(\log_{10} t_{Set})$ = -5 is achieved, (d) $P_{Switch}$ vs $|V_{Set}|$ for 50 neurons with 5% device-to-device variations, (e) Evolution of energy in the presence of device-to-device variations compared to ideal case (no variations), (f) Error% in the minimum settling energy with respect to the ideal settling energy

**Conclusion:**

In this work, the utility of PCMO RRAMs as an enabler for large scale stochastic recurrent neural networks like Boltzmann Machines is proposed. The parameters affecting the Set-time stochasticity are identified. With HRS and $V_{Set}$ fixed, $t_{Set}$ distribution is fully determined across cycles and time. The asymmetric nature of stochasticity between Set and Reset is highlighted. Deterministic and gradual state control in the Reset operation allows HRS controllability to enable drift-free stochasticity over many iterations. The reduced drift enables the solution of problems greater than 1000 nodes for the Max-Cut graphical optimization using Boltzmann Machines which is 20× higher than electrical-input only method of stochasticity generation. Further, HRS controllability allows tuning out of the device-to-device variability effects improving solution quality by 10× compared to a system with realistic variations. The properties of PCMO RRAM neuron as a stochastic neuron with controllable internal state makes it the choice of device for implementing stochasticity and weights in large scale Boltzmann Machines.